\documentclass[12pt,hyper,letterpaper]{JHEP3}
\usepackage{graphicx}
\usepackage{epsfig}
\usepackage{latexsym}
\usepackage{amssymb}
\usepackage{amsmath}
\usepackage{wrapfig}
\usepackage{boxedminipage}
\usepackage{setspace}
\usepackage{subfigure}
\usepackage{bbm}
\usepackage{xspace}

\DeclareSymbolFont{AMSb}{U}{msb}{m}{n}
\DeclareMathSymbol{\IN}{\mathbin}{AMSb}{"4E}
\DeclareMathSymbol{\IZ}{\mathbin}{AMSb}{"5A}
\DeclareMathSymbol{\IR}{\mathbin}{AMSb}{"52}
\DeclareMathSymbol{\Q}{\mathbin}{AMSb}{"51}
\DeclareMathSymbol{\II}{\mathbin}{AMSb}{"49}
\DeclareMathSymbol{\IC}{\mathbin}{AMSb}{"43}
\DeclareMathSymbol{\IP}{\mathbin}{AMSb}{"50}
\DeclareMathSymbol{\IH}{\mathbin}{AMSb}{"48}
\DeclareMathSymbol\IA{\mathalpha}{AMSb}{"41}
\DeclareMathSymbol\IS{\mathalpha}{AMSb}{"53}

\newcommand{\ads}[1]{\ensuremath{\mathrm{AdS}_{#1}}}

\newcommand{\eg}{\mbox{e.\,g.}\xspace}

\newcommand{\del}{\partial}

\newcommand{\lam}{\lambda}
\newcommand{\eps}{\epsilon}
\newcommand{\Gm}{\Gamma}
\newcommand{\dlt}{\delta}
\newcommand{\wt}{\tilde t}
\newcommand{\Lam}{\Lambda}

\title{ \LARGE A moving mirror in AdS space as a toy model for
  holographic thermalization}
\author{Johanna Erdmenger, Shu Lin and Thanh Hai Ngo \footnotemark[1]
\\Max-Planck-Institut f\"{u}r Physik (Werner-Heisenberg-Institut)
\\ F\"{o}hringer Ring 6, 80805 M\"{u}nchen, Germany}

\footnotetext[1]{E-mail addresses: \email{jke, slin, ngo  @mppmu.mpg.de}}

\date{\today}

\abstract{It is expected that thermalization may be described within
  gauge/gravity duality by considering time-dependent configurations
  on the gravity side of the correspondence, for instance a
  gravitational collapse of a matter configuration in Anti-de Sitter space. 
As a step towards
  the ambitious goal of describing such a configuration, 
we investigate a simple time-dependent toy model in which a
  mirror moves in the radial direction of Anti-de Sitter space. For
  this configuration, we establish a procedure for calculating
  two-point functions of scalar fluctuations, based on a WKB
  approximation. 
We test our method on two sample trajectories for the mirror, and find that the
singularity structure of the two-point functions is  in agreement with
 geometric optics.
}

\preprint{MPP-2011-6}

\begin{document}

\section{Introduction}

Gauge/gravity duality has been extremely successful at describing
strongly coupled systems by mapping them to weakly coupled gravity
theories. In particular, a strongly coupled
quantum field theory in $d$ dimensions at finite temperature in thermal
equilibrium can be described in terms of supergravity in a $(d+1)$-dimensional
background \ads{d+1} of Anti-de~Sitter black hole geometry \cite{Maldacena,Witten,GKP}. 
Recently, this technique has been proven to be particularly useful for describing properties of the strongly coupled quark-gluon plasma.

For further progress in holographically describing the quark-gluon
plasma as created in heavy-ion collisions, it is essential to consider
thermalization, i.e. the relaxation from an initial non-equilibrium
state to a final state in thermal equilibrium. Within gauge/gravity
duality, this is expected to be modeled by time-dependent
geometries. Describing time-dependent geometries is a difficult task in general and often requires heavy
use of numerics. Different avenues have been pursued so far towards
this goal:

One approach is to study the collision of gravitational shock waves in Anti-de Sitter space.
Efforts along this line include \cite{GR, AKT,GPY,LSK,Taliotis,CY}. 
Different aspects of heavy ion collisions have been studied in shock wave collision models,
such as early time dynamics \cite{GR,AKT,Taliotis}, entropy production  and
critical conditions on thermalization \cite{GPY,LSK, Mozo,
  Bagrov}. More recently,
the initial value problem for the non-linear Einstein equations 
has been solved numerically for planar gravitational shock waves
\cite{CY}. Despite the 
success of this approach, 
information other than the one-point function of the stress tensor
remains very difficult to obtain due to the complexity of the metric resulting from the
collision. For an initial attempt made in this direction, see \cite{GK}.

An alternative approach is to consider a collapsing matter distribution in Anti-de
Sitter space.   
It is natural to assume that in order to model far from equilibrium processes in quantum
field theories and the relaxation into equilibrium, the supergravity picture
should describe the dynamical process of black hole formation from some
initially regular spacetime. 
A mathematically clearly arranged setup is \eg
given by an infinitesimally thin but massive shell which collapses to form a
stable black hole as a final state.
Any process of black hole formation will lead from some initial problem to the
propagation of bulk fields which have to satisfy the infalling wave boundary
condition at the final equilibrium state. Before that, the boundary conditions
for the supergravity fields will be time dependent and determined by the
particular scenario. 

So far, the collapsing shell scenario has been considered in the quasi-static
approximation where the shell is considered to move adiabatically
\cite{Giddings, Danielsson,quasi}, which
simplifies the calculations significantly. In
these approaches, a collapsing thin shell geometry is probed by a scalar field or 
a graviton field. The dual boundary two-point correlator is calculated as a function 
of the radial position of the shell and thus describes the thermalization process. In
\cite{HLR}, Hubeny, Liu and Ragamani proposed a bulk-cone singularity conjecture,
which states that the two-point correlator becomes singular when the two boundary points
are connected by a bulk null geodesic. The application of this conjecture to a collapsing
shell model establishes a connection between 
a distinct signature for the boundary observables and 
horizon formation in the bulk.

A slightly different avenue in this context has been followed by
\cite{BhattacharyyaMinwalla, CheslerYaffe}. The authors of \cite{BhattacharyyaMinwalla}
demonstrate analytically that a weak scalar perturbation collapses to
form a  black hole.
The authors of \cite{CheslerYaffe} investigate the
gravitational collapse of energy injected into Anti-de Sitter space and the
formation of an event horizon by considering the evolution of locally
anisotropic metric perturbations initially located 
near the AdS boundary. At a late stage, the numerical results 
match with the analytical solutions based on an asymptotic expansion \cite{JP,Heller}. 
An interesting link 
between the thin shell and perturbative approaches described above has
recently been established in \cite{ads3,ads4,1012.4753}. The evolutions of entanglement
entropy in $d=2$ and $d=3$ have been studied in a Vaidya metric describing a collapsing
shell in the formation of a black hole.
In \cite{1012.4753} the equal-time two-point function, Wilson loop and entanglement entropy
are explored in various dimensions --
 Signatures of time dependence
for the chiral condensate have recently been investigated in
\cite{evans} by embedding a probe D$7$ brane into
the late time boost invariant background of \cite{JP}.

In this paper we start a program of studying gravitational collapse of a
thin matter shell beyond the quasi-static approximation, thus generalizing
\cite{quasi}. For clarifying a number of technical details in the study of
two-point functions for time-dependent gravity duals, in this paper we consider
a simple toy model for a time-dependent geometry consisting of a mirror moving
in the radial direction of AdS space. We impose Dirichlet boundary conditions
at the position of the mirror and calculate the two-point function of a scalar
field in this geometry. In the special case of the mirror moving with
constant velocity, scaling symmetry of AdS space is preserved. This allows us to
solve for the complete set of eigenmodes of the scalar wave in terms of scaling variables,
which are reminiscent of the late time scaling variable of Janik and
Peschanski \cite{JP} (see also the recent  \cite{kirsch}).
We compute the two point function based on the eigenmode decomposition and
find that the singularities of the two-point correlator are related to the physics of bouncing
light ray between the moving mirror and the AdS boundary. Thus the
singularity structure of the correlator is determined by a geometric
optics picture. Our results generalize the
 static mirror case considered in \cite{hoyos} to the time-dependent case.

We explore the geometric optics limit in more detail with a WKB analysis, which 
enables us to reduce a general PDE to an ODE and leads to a prescription
for calculating the two-point correlator for general radial
trajectories of the mirror. The final formula for the correlator is expressed as
a Mellin transform
involving the ratio of incoming and outgoing waves for each component in the
eigenmode decomposition. We test this correlator prescription using two sample trajectories
of the mirror. In the first case of a mirror moving with constant
velocity, we reproduce the geometric optics limit of the two-point
correlator found previously using scaling variables. In the second case where a ``mirror'' moves along a
spacelike geodesic, we find the singularities of the correlator are consistent with
the bulk-cone singularities conjecture \cite{HLR}. 

The paper is organized as follows. In section 2 we solve the eigenmodes of the scalar
wave equation in the presence of a mirror moving with constant velocity. In section 3, we
derive an explicit formula for the time-dependent, spatially integrated two-point correlator 
in terms of the eigenmodes. It takes the form of a Mellin transform. We test
this formula by considering its vacuum limit, where the mirror is absent. In section 4, based on a WKB analysis 
we establish a more general prescription for the two-point function which allows for
an arbitrary mirror trajectory. We test the prescription by reproducing the 
geometric optics limit of the correlator obtained previously. We also consider
a ``mirror'' moving along a spacelike geodesic and find the resulting correlator consistent
with the bulk cone singularities conjecture \cite{HLR}. Some details of the computation are moved
to the appendices.
\section{Moving Mirror in $\ads{d+1}$}
In this section, we calculate the scalar two-point function for a special trajectory of the mirror
which preserves scaling symmetry. We work with the  $\ads{d+1}$ metric in the Poincar\'e patch
\begin{equation}\label{AdSmetric}
ds^2=\frac{R^2}{z^2} \left( -dt^2+d\vec{x}^2+dz^2\right)\,,
\end{equation}
and start with the standard scalar wave equation
\begin{equation}\label{mastereq}
\frac{1}{\sqrt{-g}}\del_\mu(\sqrt{-g}g^{\mu\nu}\del_\nu)\phi=0 \,.
\end{equation}
\noindent
The usual way of solving the equation of motion is to consider a specific Fourier component 
$\phi(t,z)=e^{i\omega t}\phi(\omega,z)$ to reduce the PDE to an ODE. In other
words, the equation of motion is simplified by focusing on an eigenfunction 
of the operator $\del_t$. We know from quantum mechanics that this is possible,
since the operator $\del_t$ commutes with the Lagrangian operator
\begin{align}\label{lagrange}
{\cal L}\equiv\frac{1}{\sqrt{-g}}\del_\mu(\sqrt{-g}g^{\mu\nu}\del_\nu)
=\frac{z^2}{R^2}(-\del_t^2+\nabla^2+\del_z^2)-\frac{(d-1)z}{R^2}\del_z \,.
\end{align}
\noindent
An alternative explanation for the commutation of $\del_t$ and ${\cal L}$ comes from the fact that time translation symmetry  is an isometry of 
$\ads{d+1}$. In the presence of a moving mirror along the radial coordinate, the time 
translation symmetry will be broken by the Dirichlet boundary condition on the mirror.

Defining
${\bar t}=\frac{t}{R},\,{\bar x}=\frac{x}{R},\,{\bar z}=\frac{z}{R}$, we can write ${\cal L}$ in terms of dimensionless
coordinates ${\bar t},\,{\bar x} $ and ${\bar z}$. In the following the bar will be suppressed and we should keep in mind that physical quantities are measured
in units of the AdS radius $R$. Although the time translation symmetry is broken, 
scaling symmetry can be preserved for some special mirror trajectories. 
The scaling symmetry is generated by the operator
\begin{equation}
\label{eq:ScalingGenerator}
{\cal L}_x=x^{\mu}\del_{\mu}=t\,\del_t+{\vec x}\,{\vec\nabla}+z\del_z \,.
\end{equation}
\noindent
We can verify explicitly that $[{\cal L}, {\cal L}_x]=0$. 
For simplicity, we focus on solutions which depend on $(t,\,z)$ only. Solving the eigenvalue equation, we obtain
\begin{equation}\label{eigenfunc}
{\cal L}_x\phi=\lam\phi \Rightarrow \phi=v^{\lam/2}f(u)\,,
\end{equation}
\noindent
with $v=tz,\,u=\frac{t}{z}$ and $f$ being an arbitrary function of $u$. 
%
An obvious choice of the  mirror trajectory that preserves the scaling symmetry is $t/z=u_0$,
i.e. a mirror moving with constant velocity $1/u_0$. 
%
%
The general solution to \eqref{lagrange} is given by
\begin{align}\label{solLeg}
f(u)=~& A\, u^{-\lam/2}\left(u-1\right)^{\lam+\frac{1-d}{2}}F\left(\frac{1-d}{2},\frac{1+d}{2};\lam+\frac{3-d}{2};\frac{1-u}{2}\right)\nonumber\\
+ &B\, u^{-\lam/2}\left(u+1\right)^{\lam+\frac{1-d}{2}}F\left(\frac{1-d}{2},\frac{1+d}{2};-\lam+\frac{1+d}{2};\frac{1-u}{2}\right).
\end{align}
\noindent 
The solution $\phi(u,v)=v^{\frac{\lam}{2}}f(u)$ is analogous to the solution in 
momentum representation, $\phi(t,z)=e^{i\omega t}\phi(\omega,z)$. Here,   $\phi(u,v)$ describes the common eigenfunction of the Lagrangian ${\cal L}$ and
the scaling operator ${\cal L}_x$. We can view $\lam$ as playing the role of the frequency $\omega$.
Writing the solution in the more familiar coordinates $(t,z)$, we have
\begin{align}\label{phitz}
\phi(t,z)=~&A\, z^{\frac{d-1}{2}} (t-z)^{\lam+\frac{1-d}{2}} F\left(\frac{1-d}{2},\frac{1+d}{2};\lam+\frac{3-d}{2};\frac{1-t/z}{2}\right)\nonumber\\
+ &B\, z^{\frac{d-1}{2}} (t+z)^{\lam+\frac{1-d}{2}}F\left(\frac{1-d}{2},\frac{1+d}{2};-\lam+\frac{1+d}{2};\frac{1-t/z}{2}\right)\,.
\end{align}
\noindent
Near the boundary at $z=0$, we find that the scalar field given by \eqref{phitz} behaves as
\begin{equation}
\phi(t,z)\sim t^\lam+\cdots+z^d t^{\lam-d}+\cdots
\end{equation}
\noindent 
The two exponents correspond to non-normalizable and normalizable modes 
in $\ads{}$, respectively. In the flat limit $z\rightarrow\infty$,
i.e. in the deep interior of AdS, \eqref{phitz} becomes
\begin{equation}
\phi(t,z)\sim
z^{\frac{d-1}{2}}\left(a(\lam)\left(t+z\right)^{\lam+\frac{1-d}{2}}+b(\lam)\left(t-z\right)^{\lam+\frac{1-d}{2}}\right)
\, , \label{flatl}
\end{equation}
with $a(\lam)$ and $b(\lam)$ two $\lam$-dependent constants.
Obviously, $(t+z)^{\lam+\frac{1-d}{2}}$ and
$(t-z)^{\lam+\frac{1-d}{2}}$ in \eqref{flatl} correspond to 
outgoing and incoming waves. Comparing \eqref{flatl} to the flat limit
of ingoing and outcoming contributions to the scalar wave in momentum representation,
\begin{align}
&\phi(t,z)=z^{\frac{d}{2}}e^{i\omega t}H_{\frac{d}{2}}^{(1)}(\omega z)\propto  z^{\frac{d-1}{2}}e^{i\omega (t+z)} \,, \nonumber\\
&\phi(t,z)=z^{\frac{d}{2}}e^{i\omega t}H_{\frac{d}{2}}^{(2)}(\omega z)\propto  z^{\frac{d-1}{2}}e^{i\omega (t-z)} \,,
\end{align}
we conclude that $\lam$ should take the value $\lam=\frac{d-1}{2}+i\Lambda$ with $\Lambda$ an 
arbitrary real number. We now rewrite $\lam$ as $\lam= \lam'
-\frac{1-d}{2}$, 
 such that $\lam'$ is purely imaginary. In terms of $\lam'$,
 \eqref{phitz} becomes
\begin{align}
\label{ScalarSolz}
\phi(t,z)=~&Az^{\frac{d-1}{2}}(t-z)^{\lam'} F\left(\frac{1-d}{2},\frac{1+d}{2};\lam'+1;\frac{1-t/z}{2}\right) \,\nonumber\\
+&Bz^{\frac{d-1}{2}}(t+z)^{\lam'}  F\left(\frac{1-d}{2},\frac{1+d}{2};-\lam'+1;\frac{1-t/z}{2}\right).
\end{align}
For simplicity, we drop the prime on $\lambda$ in the subsequent.

\noindent
The Dirichlet boundary condition at $u=t/z=u_0$ fixes the ratio of $A$ and $B$,
\begin{equation}\label{ratio}
\frac{A}{B}  \left(\frac{u_0-1}{u_0+1}\right)^{\lam} F \left( \frac{1-d}{2},\frac{1+d}{2};\lam+1;\frac{1-u_0}{2} \right) =- F\left(\frac{1-d}{2},\frac{1+d}{2};-\lam+1;\frac{1-u_0}{2}\right).
\end{equation}
\section{The Two-Point Correlator}
\subsection{Derivation of the correlator}
In this section, we will use the solution of the scalar in the bulk to compute
the correlation functions of the dual operator in the boundary field theory.
We are interested in computing the correlation functions in coordinate space.
We perform the ``Fourier space'' analysis for the
transformation to $\lambda$ space in 
general $d$ dimensions instead of setting $d=4$.
We will follow \cite{Witten} for the computation of the two point correlator\footnote{Note a subtlety involved in this procedure is elaborated
in \cite{freedman}, but this does not affect our result for the massless scalar.}.
The two point
correlator of the operator dual to a massless scalar in $\ads{d+1}$ is given by
\begin{equation}
G(x,x')=\langle O(x)O(x')\rangle=\frac{\delta^2S}{\delta\phi^0(x)\delta\phi^0(x')} \, ,
\end{equation}
\noindent
where $S$ is the action of the scalar field. Using a regulator near the 
boundary $z=\eps$, the resultant action reads\footnote{We will suppress the overall normalization from the SUGRA action in this paper.}
\begin{align}
S_\eps&=\frac{1}{2}\int dzdtd^{d-1}x\sqrt{-g}g^{\mu\nu}\del_\mu\phi\del_\nu\phi \nonumber\\
&=\frac{1}{2}\int dtd^{d-1}x\frac{1}{z^{d-1}}\phi\del_z\phi\vert_\eps^{z_m(t)}
-\frac{1}{2}\int dzdtd^{d-1}x\phi\del_\mu(\sqrt{-g}g^{\mu\nu}\del_\nu\phi) \, .
\end{align}
\noindent 
The second term vanishes by the bulk equation of motion. Furthermore, by the Dirichlet
boundary condition $\phi$ vanishes
at the locus of the mirror $z_m(t)$ and we are left with
\begin{equation}\label{action}
S_{\eps}=-\frac{1}{2}\int dt\, d^{d-1}x\phi(t,z)\frac{\del_z\phi(t,z)}{z^{d-1}}\vert_{z=\eps} \, .
\end{equation}
Note that our scalar wave in the bulk has no dependence on spatial $x$. The vertex that couples the source
$\phi(t,x)$, the boundary value of $\phi(t,x,z)$, with the operator $O(t,x)$ simplifies to
\begin{equation}\label{actionSimplifying}
 \int dt\,d^{d-1}x\phi(t,x)O(t,x)=\int dt\phi(t)\int d^{d-1}xO(t,x)\, .
\end{equation}
\noindent 
The functional derivative of the action with respect to $\phi(t)$ gives
\begin{equation}\label{correlator}
\langle \int d^{d-1}xO(t,x)\int d^{d-1}x'O(t',x')\rangle
=-\lim_{z\rightarrow 0}\frac{\delta\del_z\phi(t,z)}{z^{d-1}\delta\phi(t',z)}\text{Vol} \, ,
\end{equation}
\noindent
with $\text{Vol}=\int d^{d-1}x$ being the spatial volume, since we assume a spatially infinite
 mirror that does not break translational invariance in spatial dimensions. The causal nature
of (\ref{correlator}) will be specified in each explicit example later and discussed in section 4.
%
\\
\\
\noindent 
Expanding \eqref{ScalarSolz} near $z=0$, the solution to \eqref{mastereq} takes the form
\begin{align}\label{expansion}
\phi_{\lam}(t,z)
=  K(d, \lam, A, B) ~ t^{\lam+\frac{d-1}{2}} +\cdots 
+ L(d, \lam, A, B)~t^{\lam-\frac{d+1}{2}} z^d+\cdots ,
\end{align}
with
\begin{align}
\label{Eq:DefineKL}
K(d, \lam, A, B)&=\left(\frac{1}{2}\right)^{\frac{-1+d}{2}}\frac{\Gm(d)}{\Gm(\frac{1+d}{2})}\left[\frac{\Gm(1+\lam)}{\Gm(\frac{1+d}{2}+\lam)}A +\frac{\Gm(1-\lam)}{\Gm(\frac{1+d}{2}-\lam)}B \right]\nonumber \,,\\
L(d, \lam, A, B)&=\left(\frac{1}{2}\right)^{\frac{-1-d}{2}}\frac{\Gm(-d)}{\Gm(\frac{1-d}{2})}\left[\frac{\Gm(1+\lam)}{\Gm(\frac{1-d}{2}+\lam)}A +\frac{\Gm(1-\lam)}{\Gm(\frac{1-d}{2}-\lam)}B \right] \,,
\end{align}
and the $\cdots$ denote terms of the form $W(d,\lam) z^{d'}$ with integer $d'\neq d$. We do not write down $W(d,\lam) z^{d'}$ explicitly, since they do not contribute to the two-point correlation functions \eqref{correlator}. Using the eigenfunction (\ref{solLeg}) as a basis set, we can express an arbitrary wave
$\phi(t,z)$ as a superposition of $\phi_\lam(t,z)$, namely
\begin{eqnarray}\label{coeff}
\phi(t,z)=\int\phi_\lam(t,z)g(\lam)d\lam&=&\int\phi^0_\lam(t)g(\lam)d\lam+\cdots+z^d \int\phi^d_\lam(t)g(\lam)d\lam +\cdots \nonumber\\
&\equiv&\phi^0(t)\cdots+z^d\,\phi^d(t)+\cdots \, .
\end{eqnarray}
Here $g(\lam)$ describes the weighting function for the component $\phi_\lam(t,z)$ with eigenvalue $\lam$. In the following we will look for an explicit expression for the weighting function $g(\lam)$, then using that result we will be able to write $\phi^d(t)$ as a functional of $\phi^0(t)$. This is the crucial step to determine the two point-correlator coming from inserting (\ref{coeff}) in (\ref{correlator}), giving
\begin{eqnarray}\label{funcderv}
\langle\! \int d^{d-1}xO(t,x)O(t',0)\rangle=-d\frac{\delta \phi^d(t)}{\delta \phi^0(t')}\,,
\end{eqnarray}
where contact terms from $\cdots$ in (\ref{coeff}) are excluded.

At the moment we do not care about the causal nature of the correlator (\ref{funcderv}), but later we will discuss it in subsection \ref{CorrLimits}.
The relevant explicit decomposition of $\phi^0(t)$ and $\phi^d(t)$ defined in (\ref{coeff}) are obtained
from (\ref{expansion}) as%
\begin{subequations}
\begin{align}
\phi^0(t)&=\int_{-i\infty}^{+i\infty} K(d, \lam, A, B) ~ t^{\lam+\frac{d-1}{2}} ~g(\lam)\, d\lam \,,\label{phi0}\\
\phi^d(t)&=\int_{-i\infty}^{+i\infty} L(d, \lam, A, B) ~t^{\lam-\frac{d+1}{2}}~g(\lam)\, d\lam \,.\label{phid} 
\end{align}
\end{subequations}
Defining $\bar\phi^0(t)=\phi^0(\frac{1}{t})$, we identify 
\eqref{phi0} as the inverse Mellin transform. This observation allows us to
invert (\ref{phi0}) using the Mellin transform and obtain $g(\lam)$ via the relation
\begin{equation}\label{glam}
K(d, \lam, A, B) ~g(\lam)=\frac{1}{2\pi i}\int_0^\infty\bar\phi^0(t)t^{\lam+\frac{d-3}{2}}dt   =  \frac{1}{2\pi i}\int_0^\infty\phi^0(t')t'{}^{-\lam-\frac{d+1}{2}}dt'    \,, 
\end{equation}  
\noindent
where in the intermediate step, a change of variable $t=1/t'$ is used. Plugging \eqref{glam} to \eqref{phid}, we obtain
\begin{equation}
\phi^d(t)=\frac{1}{2\pi i}\int_{-i\infty}^{+i\infty}\frac{L(d, \lam, A, B)}{K(d, \lam, A, B)}~
 t^{\lam} \int_0^{\infty} t'{}^{-\lam}(tt')^{-\frac{d+1}{2}}\phi^0(t')~dt' d\lam\,.
\end{equation}
Using \eqref{funcderv}, we end up with the  correlator in the integral representation 
\begin{equation}\label{lam_int}
\langle\! \int\! d^{d-1}xO(t,x)O(t',0)\rangle=-\frac{d }{2\pi i}\int_{-i\infty}^{+i\infty}\frac{L(d, \lam, A, B)}{K(d, \lam, A, B)}\times \frac{ t^{\lam}t'{}^{-\lam}}{(tt')^{\frac{d+1}{2}}}d\lam \,,
\end{equation}
with $K(d, \lam, A, B)$ and $L(d, \lam, A, B)$ defined in \eqref{Eq:DefineKL} and $A,\,B$ satisfying (\ref{ratio}).

Before we proceed to the evaluation of this expression, 
several comments on the correlator are in order: i) As is common in
Minkowski signature, the correlator obtained from (\ref{lam_int}) will
depend on the specific wave we use in the bulk. In the
mirror geometry, the causal structure of the correlator is in general
complicated. In a certain limit, it should reduce to the retarded ($B=0$) or 
advanced ($A=0$) correlator. We will see later that the limiting correlator does agree
with those obtained in momentum space representation;
ii) We have derived the correlator for $u_0>1$, which is defined for
$t,t'>0$. Actually most formulae are equally true for $u_0<-1$, corresponding
to $t,t'<0$. We will however, focus on the case $u_0>1$ for definiteness
in what follows;
iii) Our moving mirror does not introduce any 
dissipation to the background, the correlator should therefore be real;
iv) The correlator should be finite. We note the possible divergent factor 
$\frac{\Gm(-d)}{\Gm(\frac{1-d}{2})}$ when $d=2,4,6\cdots$. This potential
pole should be cancelled out by the $\lam$-integral.\footnote{In those dimensions,
the scalar wave contains a logarithmically divergent term near the boundary, 
which is encoded in the definition of the Hypergeometric function.}
%
\subsection{Different limits of the correlator}
\label{CorrLimits}
Before evaluating the integral (\ref{lam_int}), let us look at the limit
$B=0$ and $A=0$. These correspond to the incoming and outgoing 
waves, respectively. 
We should expect (\ref{lam_int}) to give retarded and advanced correlators
from experience in the momentum space representation. We show in the following that
this is also true in the $\lam$ representation, which will serve as a nontrivial
check of our prescription. 
For definiteness, we choose $t,t'>0$. At $B=0$, the
$\lam$ dependent part of the integrand simplifies to 
$\frac{\Gm(\lam+\frac{1+d}{2})}{\Gm(\lam+\frac{1-d}{2})}\left(\frac{t}{t'}\right)^\lam$. The poles in the
complex $\lam$ plane are at $\lam=-n-\frac{1+d}{2}$ with integer $n\ge 0$. 
The integral is only nonvanishing
when the integration contour is closed counter-clockwise, i.e. when $t>t'$.
Summing over  residues, we obtain for the correlator defined in 
\eqref{funcderv}
\begin{eqnarray}\label{retarded}
\frac{\delta\phi^d(t)}{\delta\phi^0(t')}=\theta(t-t')\frac{2^d}{(t-t')^{d+1}}\frac{\Gm(\frac{1+d}{2})}
{\Gm(\frac{1-d}{2})\Gm(d)} \,.
\end{eqnarray}
For $A=0$, the situation is quite similar. 
The $\lam$-dependent integrand simplifies to 
$\frac{\Gm(\frac{1+d}{2}-\lam)}{\Gm(\frac{1-d}{2}-\lam)}\left(\frac{t}{t'}\right)^\lam$, with the poles located
at $\lam=n+\frac{1+d}{2}$ with integer $n\ge 0$. Physically we require $d\ge 1$, and as a result
the integral is only nonvanishing when the integration contour is closed
clockwise, i.e. when $t<t'$. Summing over the residues, we obtain
\begin{eqnarray}\label{advanced}
\frac{\delta\phi^d(t)}{\delta\phi^0(t')}=\theta(t'-t)\frac{2^d}{(t'-t)^{d+1}}\frac{\Gm(\frac{1+d}{2})}
{\Gm(\frac{1-d}{2})\Gm(d)} \,
\end{eqnarray}
\noindent for the correlator.
Let us now compare this result with the standard momentum and frequency
representation, in which the incoming wave is given by
\begin{eqnarray}
\phi_{in}=
\left\{\begin{array}{l@{\quad\quad}l}
z^{d/2}H_{\frac{d}{2}}^{(2)}(\omega z)& \omega>0\\
z^{d/2}H_{\frac{d}{2}}^{(1)}(-\omega z)& \omega<0  ,
\end{array}
\right.
\end{eqnarray}
which gives rise to
\begin{eqnarray}
\frac{\delta\phi^d(\omega)}{\delta\phi^0(\omega)}=
\left\{\begin{array}{l@{\quad\quad}l}
-e^{\frac{i\pi d}{2}}\frac{\Gm(1-\frac{d}{2})}{\Gm(1+\frac{d}{2})}(\frac{\omega}{2})^d& \omega>0\\
-e^{\frac{-i\pi d}{2}}\frac{\Gm(1-\frac{d}{2})}{\Gm(1+\frac{d}{2})}(\frac{-\omega}{2})^d& \omega<0 \,.
\end{array}
\right.
\end{eqnarray}
Fourier transforming the above back to coordinate space, we obtain
\begin{eqnarray}\label{omega}
\frac{\delta\phi^d(t)}{\delta\phi^0(t')}=-\theta(t-t')\frac{\Gm(d+1)}{(t-t')^{d+1}} 
\frac{\Gm(1-\frac{d}{2})}{\Gm(1+\frac{d}{2})}\frac{\cos\left(\pi(d+\frac{1}{2})\right)}{2^d~\pi}\,.
\end{eqnarray}
After writing the $\cos$-term as a product of two $\Gamma$-functions and using some relations between the $\Gm$-functions, it can be shown that
(\ref{omega}) and (\ref{retarded}) are indeed identical, demonstrating
the equivalence of using the $\lambda$ or the $\omega$
representation. 

Keen readers may have noticed that our correlator 
vanishes for odd $d$. This is true in the domain of time we are
interested in, i.e. for  $t\ne t'$.
In Appendix B, we present explicit examples of spatially integrated correlators for $d=3$ and $4$
starting with a general formula for unintegrated correlators in CFT\footnote{At $t=t'$, there is
a non-analyticity resulting from the lightcone non-analyticity of the unintegrated correlators.}. These examples clearly display the subtle difference between the calculation
in odd and even dimensions.

A similar analysis shows that the advanced
correlator obtained from the momentum representation also agrees with 
(\ref{advanced}). This boosts our confidence in the $\lam$ representation
of the correlator, and we will extract the time-dependent correlator in the
moving mirror background using this representation.

We focus on the UV part of the
correlator following the work by Amado and Hoyos
\cite{hoyos}. Specifically, these authors
 show that 
the UV part of the correlator, i.e. the part obtained by considering
only frequencies with $|\omega| \gg 1$,  has an equivalent description in terms of
 geometric optics in $\ads{}$. Moreover, they found that the singularities of
the correlator correspond to the time when the light ray bouncing between the
$\ads{}$ boundary and the mirror hits the boundary. This is a special case of
the bulk-cone singularities conjecture by Hubeny, Liu and Rangamani \cite{HLR}.
The latter, originally formulated in global AdS space, states that the singularities occur
when the two boundary points are connected by a bulk null geodesic,
i.e. by a light ray
trajectory. In the Poincar\'e patch, the light ray will not return to the boundary without 
being reflected at the mirror.
It should be stressed that while the authors of \cite{hoyos} use a static mirror
in $\ads{}$ which introduces an explicit scale
to the boundary CFT, our mirror moves in such a way that scale invariance of the
boundary CFT is still preserved. Therefore, the UV limit of our case amounts to
summing over all residues in the complex $\lam$ plane, with $|\lam|\gg 1$.
As we will see, the geometric picture is robust in our moving mirror geometry. 

The UV part of the
correlator is evaluated in appendix A. As explained there in detail, the causal nature
 of the correlator is related to the chosen integration contour of
 $\lam$. In particular, in the UV limit, which amounts in particular to using the relations \eqref{App:A}, the 
retarded correlator is given by 
\begin{align}\label{mirror_u0}
\big\langle\! d^{d-1}x~O(t,x)O(t',0)\big\rangle_R =&-\theta(t-t')\sum_{+,-}(tt')^{\frac{-d-1}{2}}
d\left(\frac{2}{a}\right)^d\frac{1-e^{\mp i\pi d}}{a}\frac{\Gm(-d)\Gm(\frac{1+d}{2})}{\Gm(\frac{1-d}{2})\Gm(d)} \nonumber\\
\times& e^{\pm \frac{i\pi(d-1)}{2}(\frac{b}{a}-c)}
\bigg[\frac{d!}{(-c)^{d+1}(\pm i2\pi)}-\sum_{r=0}^\infty\frac{B_{d+r+1}(\frac{d-1}{4})c^r(\pm i2\pi)^{d+r}}{r!(d+r+1)}\bigg] \nonumber\\
\equiv&-\theta(t-t')\frac{d}{a^{d+1}}(tt')^{\frac{-d-1}{2}}g\left(\frac{\pi b}{a}\right)\,,
\end{align}
\noindent 
with $a=\ln\frac{u_0+1}{u_0-1}$ and $b=\ln\frac{t}{t'}$. $B_n(x)$ are the Bernoulli polynomials and 
\begin{equation}\label{def_c}
e^{i2\pi c}=e^{\frac{i2\pi b}{a}},\quad |c|<\frac{1}{2}.
\end{equation}
\EPSFIGURE{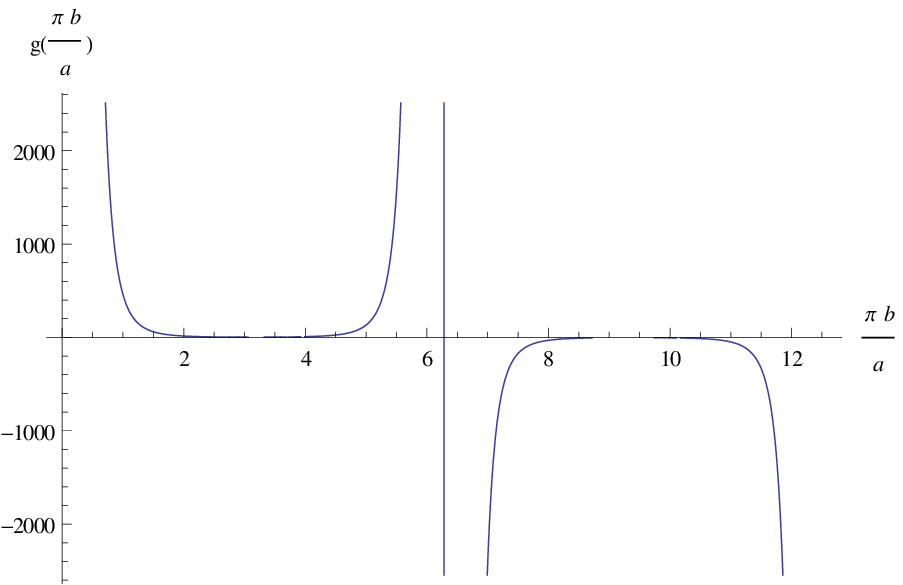,width=0.6\textwidth}{
The correlator contribution $g\left(\frac{\pi b}{a}\right)$ as defined in (\ref{mirror_u0}) versus 
$\frac{\pi b}{a}$ at $d=4$. $g\left(\frac{\pi b}{a}\right)$  has a period
of $4\pi$. In every odd interval $(2n\pi,(2n+1)\pi)$, the correlator 
is positive, while in every even interval $((2n+1)\pi,(2n+2)\pi)$, 
the correlator flips sign. \label{oscillate}}
\EPSFIGURE{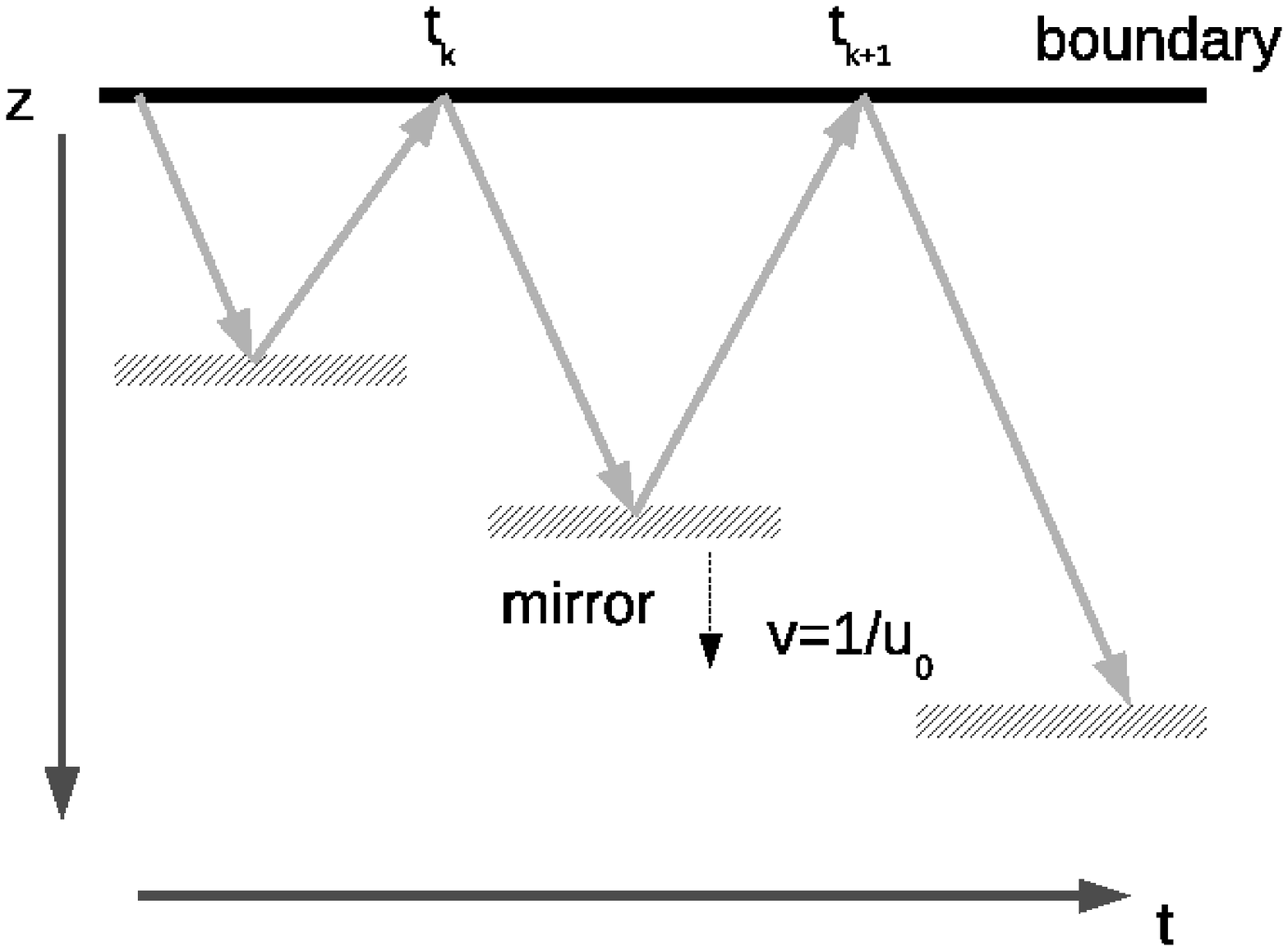,width=0.6\textwidth}
{A schematic picture of a moving mirror in the $\ads{}$ background, and the trajectory
of a bouncing light ray. The correlator we obtain
displays a
singularity structure in agreement with this geometric optics picture. \label{bouncing}}
Note that the square bracket is only a function of $c$, which according to 
(\ref{def_c}) is a periodic function with period $1$. As $c\rightarrow 0$,
the first term in the square bracket is singular while the second term
is regular. The correlator diverges when $b/a=n$, with $n \in\text{N}$. 
Another interesting observation is that for odd
dimensions ($d=1, 3, \cdots$), the correlator  vanishes identically due
to the $d$-dependent prefactor, which is 
a remnant of the behavior of the vacuum correlator.

The periodic divergence of the correlator in the moving mirror geometry is
consistent with the expectation from the geometric optics limit proposed
in \cite{hoyos}: Suppose the mirror starts moving with velocity
$1/u_0$ at $t=0$ from the
$\ads{}$ boundary. If at $t_k$ the light ray reaches the boundary,
it is easy to see that the next time the light ray hits the boundary
will be at
$t_{k+1}=\frac{u_0+1}{u_0-1}t_k$, thus 
$\ln\frac{t_{k+1}}{t_k}=\ln\frac{u_0+1}{u_0-1}$, which amounts to
$b=a$. Indeed at
$b/a=n$ when $b$ is an integer multiple of $a$, we observe singularities in the correlator.
A plot of the correlator is included in
Fig.\ref{oscillate}. The structure of the correlator deserves some explanations: The function 
$g\left(\frac{\pi b}{a}\right)$ has a period of $4\pi$, which is due to the oscillatory factor 
$e^{\pm \frac{i\pi(d-1)}{2}(\frac{b}{a}-c)}=e^{\pm \frac{i\pi(d-1)}{2}n}$.
The period of the singularities in the correlator is $2\pi$ because there are two
singularities in one period of $g\left(\frac{\pi b}{a}\right)$.
From (\ref{mirror_u0}), one might expect the period to be $\pi$, however
 half of the singularities
vanish as we sum over the plus and minus contributions in (\ref{mirror_u0}). 
This is an artifact of choosing the retarded correlator. The Feynman correlator e.g.
will pick either a plus or a minus sign, displaying the other half of the singularities explicitly.
We have also included a view of a bouncing light ray in the presence of
the moving mirror in Fig. \ref{bouncing}. This figure shows 
schematically that
the time separation of the singularities increases  with time as the mirror moves
further and further away from the boundary.
\section{Moving Mirror in the Limit of Geometric Optics}
\subsection{The WKB approximation and the limit of geometric optics}
As we have derived in the previous section, the two point correlator in the UV limit contains singularities
when the times are related by a geometric optics path. In this
section, we make this connection more precise. In particular, we
develop a prescription for calculating the two-point function  
for general trajectories of the mirror. As a  test of our prescription, we reproduce
the result of the two-point function obtained in the previous section.

The geometric optics limit can be described by a WKB solution for the
scalar wave in the $\ads{}$ background.
Let us write the scalar wave as $\phi=Ae^{i\theta/\eps}$, with $A$ and $\theta$ being the
amplitude and the phase of the wave. The essence of the WKB approximation is that the phase $\theta$
varies much faster than the amplitude $A$. We plug $\phi=Ae^{i\theta/\eps}$ to the equation of motion 
$\square\phi=0$ and perform a series expansion in $\eps$. The equation of motion for $A$ and $\theta$
are given by the leading order and next-to-leading order terms\footnote{We focus only on spatially homogeneous solutions and discard derivatives with respect to $\vec{x}$.}
\begin{eqnarray}
&&-\dot{\theta}^2+\theta'{}^2=0 \,, \\
&&-2\dot{A}\dot{\theta}+2A'\theta'+(-\ddot{\theta}+\theta'')A-\frac{d-1}{z}A\theta'=0 \,,
\end{eqnarray}
where the dot and prime denote derivatives with respect to $t$ and $z$, respectively.
The first equation can be solved for $\theta=\theta_\pm(t\pm z)$. They will be used to
eliminate the bracket in the second equation. The latter is 
solved by $A=z^{\frac{d-1}{2}}A_\pm(t\pm z)$. The positive and
negative sign solutions have obvious 
identifications as outgoing and incoming waves. They form two linearly independent solutions
to the EoM. Therefore, we split the solution of $ \square \phi = 0$ into
\begin{align} \label{Fpm}
\phi(t,z)=z^{\frac{d-1}{2}}\left(A_+(t+z)e^{\frac{i\theta_+(t+z)}{\eps}}+A_-(t-z)e^{\frac{i\theta_-(t-z)}{\eps}}\right)\equiv z^{\frac{d-1}{2}}\left(F_+(t+z)+F_-(t-z)\right) \,.
\end{align}
The WKB solution breaks down near the singularities of the original equation of motion $\square\phi=0$.
In the case of the $\ads{}$ background, the only singularity is the
AdS boundary $z=0$. 
Near the AdS
boundary, a general scalar wave can be written as a superposition of eigenmodes,
\begin{align}\label{super}
\phi(t,z)=\int \bigg[&g_-(\lam) z^{\frac{d-1}{2}} (t-z)^{\lam} F\left(\frac{1-d}{2},\frac{1+d}{2};\lam+1;\frac{1-t/z}{2}\right) \nonumber\\
+&g_+(\lam)z^{\frac{d-1}{2}}(t+z)^{\lam}F\left(\frac{1-d}{2},\frac{1+d}{2};-\lam+1;\frac{1-t/z}{2}\right)\bigg]d\lam \,.
\end{align}
where $g_\pm(\lam)$ are the weighting functions for the incoming and
outgoing components with eigenvalue $\lam$, respectively. 
In the UV limit $|\lam|\rightarrow\infty$, (\ref{super}) can be nicely matched with the WKB solution.
Since $\lim_{\gamma\rightarrow\infty}F(\alpha,\beta;\gamma;z)=1$, equation (\ref{super}) simplifies to
\begin{eqnarray}
\phi(t, z)=z^{\frac{d-1}{2}}\int\bigg[g_-(\lam)(t-z)^{\lam}+
g_+(\lam)(t+z)^{\lam}\bigg]d\lam \,,
\end{eqnarray}
and we identify
\begin{eqnarray}
F_+(t+z)&=&\int g_+(\lam)(t+z)^{\lam}d\lam \,, \\ 
F_-(t-z)&=&\int g_-(\lam)(t-z)^{\lam}d\lam \,,
\label{FpFm}
\end{eqnarray}
with $F_+$ and $F_-$ defined in \eqref{Fpm}.

We now aim for an expression for the correlator in 
a more general time-dependent situation, where the mirror follows a 
general, not necessarily scale invariant trajectory. Before proceeding to the derivation
of such an expression, let us take a moment to look at the setting of the 
problem. Since we are interested
in a time-dependent scalar wave in the bulk, which amounts to solving a 2D
PDE (in the absence of $\vec{x}$ dependence), we need to specify both boundary
conditions and initial conditions in order to have a unique solution. The
boundary conditions are provided by the source at the $\ads{}$ boundary
and the Dirichlet boundary condition at the mirror. There is no further initial condition.

The mirror with trajectory $z=t/u_0$ is special in the sense that it allows us
to study the ``steady wave'' (analogous to a planar wave in the AdS
 background), which does not require an initial condition.
This particular trajectory does not couple the eigenmodes with different $\lam$ and allows us to determine a ratio $\frac{g_+(\lam)}{g_-(\lam)}$ for every $\lam$.
In terms of the boundary field theory, the missing
initial condition is  encoded in the state on which the
operator $O(t,x)$ acts.

 In non-equilibrium field theory, the correlator
should be studied using the in-in contour \cite{noneq}, which is composed of one forward
and one backward contour in the complex time plane. This was formulated holographically in
 real-time gauge/gravity duality by Skenderis and van Rees \cite{vanrees}. 
The state should be prepared by inserting
sources in the Euclidean segments of the in-in contour. One of the essential points
of \cite{vanrees} is that the bulk solution should be completely fixed by matching the
Lorentzian segments to the Euclidean ones. The matching effectively provides the
initial condition. Our
mirror can be viewed as an effective source which creates the state.  Despite 
the ambiguity related to the missing initial condition as described
below \eqref{FpFm}, we will see that taking the UV limit as before actually
allows us to obtain the wave solution without specifying an initial condition.
In the UV limit  $|\lam|\rightarrow\infty$, the solution in the bulk only depends on $t\pm z$, as defined in \eqref{Fpm}, thus reducing
the PDE to an ODE. 

To solve the wave equation, we start with the
Dirichlet boundary condition on the trajectory of the mirror which
implies that the wave vanishes there,
\begin{equation}\label{gr_zm}
\int\bigg[g_-(\lam)\left(t-z_m(t)\right)^{\lam}+g_+(\lam)(t+z_m(t))^{\lam}\bigg]d\lam=0 \,.
\end{equation}
Here and below, we suppress the integration bounds, with the understanding that the $\lam$
integral runs always from $-i\infty$ to $+i\infty$,
while the $t$ integral runs from $0$ to $\infty$.
This is to be combined with the asymptotic behavior near $z=0$ similar to \eqref{expansion}, where the unique weighting function $g(\lam)$ in  \eqref{coeff} is replaced by the more general weighting functions $g_\pm(\lam)$ as described in \eqref{super},
\begin{eqnarray}
\label{PhiZero}
\phi^0(t)&=&\left(\frac{1}{2}\right)^{\frac{d-1}{2}} \frac{\Gm(d)}{\Gm(\frac{1+d}{2})}\int \left( K_+\left( \lam\right)g_-(\lam)+K_-\left( \lam\right)g_+(\lam)\right) t^{\lam+\frac{d-1}{2}} d\lam \,,
\end{eqnarray}
with
\begin{align} \label{K}
K_+(\lam)=\frac{\Gm(1+\lam)}{\Gm(\frac{1+d}{2}+\lam)}\,,\quad K_-(\lam)=\frac{\Gm(1-\lam)}{\Gm(\frac{1+d}{2}-\lam)}\,.
\end{align}
For the sake of clarity, we introduce a shorthand notation $\tilde \phi(\lam)$ defined as
\begin{align}
\label{bdry_s}
\tilde\phi(\lam) \equiv  K_+( \lam)g_-(\lam)+K_-( \lam)g_+(\lam) \,,
\end{align}
hence equation \eqref{PhiZero} can be written as 
\begin{equation} 
\label{PhiZeroSimplified}
 \phi^0(t)=\left(\frac{1}{2}\right)^{\frac{d-1}{2}} \frac{\Gm(d)}{\Gm(\frac{1+d}{2})}\int\tilde\phi(\lam)~ t^{\lam+\frac{d-1}{2}} d\lam.
 \end{equation} 
Next we will use equations (\ref{gr_zm}) and the definition in (\ref{bdry_s}) to write $g_\pm(\lam)$ in terms of  $\tilde\phi(\lam)$. Later we will use the Mellin transform of \eqref{PhiZeroSimplified} to express $g_\pm(\lam)$ as functions $\phi^0(t)$. Then using these results, we can determine the coefficient function  $\phi^d(t)$ as an expression of  $\phi^0(t)$ and compute the two-point correlation functions. First using (\ref{bdry_s}) for rewriting (\ref{gr_zm}), we obtain two relations
\begin{eqnarray}\label{gpm}
 \int \left(K_+(\lam)(t+z_m)^\lam-K_-(\lam)(t-z_m)^\lam\right)\frac{g_-(\lam)}{K_-(\lam)}d\lam&=&\int\frac{\tilde\phi(\lam)}{K_-( \lam)}(t+z_m)^\lam d\lam \,,~~~\\
\int\left(K_-( \lam)(t-z_m)^\lam-K_+(\lam)(t+z_m)^\lam\right)\frac{g_+(\lam)}{K_+(\lam)}d\lam&=&\int\frac{\tilde\phi(\lam)}{K_+(\lam)}(t-z_m)^\lam d\lam \,.
\label{gpm2}
\end{eqnarray}
Defining ${\cal D}_F(\lam,t)$ as the inverse of the time-dependent part on the LHS of \eqref{gpm} with respect to integration over $t$, i.e.
\begin{align}
\label{DefF}
K_+( \lam')(t+z_m(t))^{\lam'}-K_-(\lam')(t-z_m(t))^{\lam'}&\equiv F(\lam',t), \nonumber \\
\int \! dt\,{\cal D}_F(\lam,t)F(\lam',t)&=\frac{1}{i}\dlt(\frac{\lam-\lam'}{i}) \,.
\end{align}
The insertion of the $i$'s is to remind us that $\lam$ and $\lam'$ are purely imaginary. 
The solutions for \eqref{gpm} and \eqref{gpm2} can be formally written as
\begin{eqnarray}\label{Fredholm_sol}
 \frac{g_-(\lam)}{K_-(\lam)}&=&\int dt d\lam'~{\cal D}_F(\lam,t)\frac{\tilde\phi(\lam')}{K_-( \lam')}(t+z_m)^{\lam'}\,,\nonumber \\
 \frac{g_+(\lam)}{K_+( \lam)}&=&-\int dt d\lam'~{\cal D}_F(\lam,t)\frac{\tilde\phi(\lam')}{K_+(\lam')}(t-z_m)^{\lam'} \,,
\end{eqnarray}
respectively. %
%
%
The next step is to determine the near boundary coefficient function  $\phi^d(t)$ in terms of  $\phi^0(t)$. Similarly to \eqref{PhiZero}, $\phi^d(t)$ is given by
\begin{eqnarray}
\label{PhiD}
\phi^d(t)&=&\left(\frac{1}{2}\right)^{\frac{-1-d}{2}}\frac{\Gm(-d)}{\Gm(\frac{1-d}{2})} \int \left( L_+\left( \lam\right)g_-(\lam)+L_-\left( \lam\right)g_+(\lam)\right) t^{\lam-\frac{d+1}{2}} d\lam \,,
\end{eqnarray}
with 
\begin{equation}
\label{LpLm}
L_+(\lam)=\frac{\Gm(1+\lam)}{\Gm(\frac{1-d}{2}+\lam)} \,,
\quad
L_-(\lam)= \frac{\Gm(1-\lam)}{\Gm(\frac{1-d}{2}-\lam)}\,.
\end{equation}
After using the Mellin transform of \eqref{PhiZeroSimplified} 
\begin{equation}
\label{MellinPhiZero}
\tilde\phi(\lam')=\frac{1}{2\pi i}\frac{\Gm(\frac{1+d}{2})}{\Gm(d)}\left(\frac{1}{2}\right)^{\frac{-d+1}{2}}\int dt'\phi^0(t')t'{}^{-\lam'-\frac{d+1}{2}}\,
\end{equation}
to replace $\tilde\phi(\lam')$ by $\phi^0(t')$ in \eqref{Fredholm_sol}
and then inserting (\ref{Fredholm_sol}) into (\ref{PhiD}), we obtain 
\begin{align}\label{coeff_d}
\phi^d(t)=\frac{2^d}{2\pi i} \frac{\Gm(-d)\Gm(\frac{1+d}{2})}{\Gm(\frac{1-d}{2})\Gm(d)}&\int d\lam d\lam' dt'' ~
N(\lam,\lam',t'',z_m(t''))
\int dt'~ \phi^0(t')   \frac{ t^{\lam}t'{}^{-\lam'}}{(tt')^{\frac{d+1}{2}}}\,, \\
N(\lam,\lam',t'',z_m)\equiv {\cal D}_F(\lam,t'')&\left(L_+(\lam)K_-(\lam)\frac{(t''+z_m)^{\lam'}}{K_-(\lam')} 
-L_-(\lam)K_+(\lam)\frac{(t''-z_m)^{\lam'}}{K_+(\lam')}\right) \,. \nonumber
\end{align}
Using \eqref{funcderv}, we end up with the  correlator in the integral representation, 
\begin{align}\label{Gttp}
\langle\! \int\! d^{d-1}xO(t,x)O(t',0)\rangle=\frac{-2^d d}{2\pi i}\frac{\Gm(-d)\Gm(\frac{1+d}{2})}{\Gm(\frac{1-d}{2})\Gm(d)}\int d\lam d\lam' dt'' 
N(\lam,\lam',t'',z_m(t''))  \frac{ t^{\lam}t'{}^{-\lam'}}{(tt')^{\frac{d+1}{2}}} \,.
\end{align}
This is our main result.
The main difference between \eqref{Gttp}  and \eqref{lam_int} are the
integrations over $\lam'$ and $t''$  in \eqref{Gttp}  which encode the
motion of the mirror for non-constant velocity. In the
following we will see that \eqref{Gttp} and  \eqref{lam_int} indeed
give the same results if the mirror moves with constant velocity along
the radial coordinate. For this purpose, we apply the result \eqref{Gttp} to the case of a mirror moving with constant 
velocity $t/z_m=u_0$. Then $F(\lam,t)$ and its inverse ${\cal D}_F(\lam,t)$ defined in \eqref{DefF}
are given by
\begin{eqnarray}
F(\lam,t)&=&\left(K_+(\lam)\left(1+\frac{1}{u_0}\right)^\lam-K_-(\lam)\left(1-\frac{1}{u_0}\right)^\lam\right)t^\lam \,, \\
{\cal D}_F(\lam,t)&=&\frac{1}{2\pi i}\frac{t^{-\lam-1}}{K_+(\lam)(1+\frac{1}{u_0})^\lam-K_-(\lam)(1-\frac{1}{u_0})^\lam} \,.
\end{eqnarray}

Plugging the above expression to (\ref{Gttp}), we see the integrals of $t''$ and $\lam'$ are 
trivial. The final integral of $\lam$ is identical to (\ref{lam_int}) up to
setting the hypergeometric 
function to $1$, which is precisely the UV limit which we used in appendix A for the
evaluation of the correlator. As a result, we are bound to reproduce (\ref{mirror_u0}).

\subsection{The Correlator in the presence of a non-constantly moving ``mirror''}

In this section, we will illustrate the procedure for obtaining the correlator in the presence of a mirror
moving with non-constant velocity. We will see that the results are again consistent with expectations 
from geometric optics. The procedure derived in the previous subsection works for any trajectory 
of the mirror, but in practice this is hard to deal with for
complicated trajectories, since it involves the inversion of the integral operator in order to 
obtain ${\cal D}_F(\lam,t)$ from $F(\lam,t)$. In principle, this is
possible to calculate by applying the Fredholm
theory \cite{tricomi}. However, this can be extremely complicated. We aim at finding some
special trajectories which can lead to significant simplifications of the inversion procedure. We will
see that it is possible at least in the UV limit. 

Suppose we propose the trajectory with $t-z_m=\frac{1}{t+z_m}$, or equivalently $z_m=\sqrt{t^2-1}$, defined
for $t>1$. This is a spacelike trajectory, thus cannot be associated with a physical mirror. We should
understand it as providing a boundary condition in the bulk. Since this is a spacelike trajectory, any
light emitted from the boundary into the bulk after $t=1$ will 
not catch the ``mirror'' and no reflection is expected. Therefore, we should not see any
singularities in the two point correlator according to the geometric optics picture. We will confirm this
by explicitly computing the correlator in the UV limit.\\
\\ \noindent
Note that $F(\lam,t)=K_+(\lam)(t+z_m)^\lam-K_-(\lam)(t-z_m)^\lam$  as defined in \eqref{DefF} and $K_+(\lam)=K_-(-\lam)$. With the proposed trajectory, we have
\begin{equation}
F(\lam,t)=K_+(\lam)(t+z_m)^\lam-K_+(-\lam)(t+z_m)^{-\lam}=-F(-\lam,t) \,.
\end{equation}
In the limit $\lam\rightarrow\infty$, $K_+(\lam)=\frac{\Gm(\lam+1)}{\Gm(\lam+\frac{1+d}{2})}\sim\lam^{\frac{1-d}{2}}$, where the argument is fixed by $|\text{arg}(\lam)|<\pi$. Writing $\lam=i\Lam$,  we have the
explicit expression for $F(\lam,t)$
\begin{eqnarray}
F(\lam,t)&=&\text{sgn}(\Lam)\,|\Lam|^{\frac{1-d}{2}}\,2i\,\sin\left(|\Lam|\,\wt+\frac{\pi(1-d)}{4}\right) \\ \nonumber
&=&\text{sgn}(\Lam)|\Lam\,|^{\frac{1-d}{2}}\,2i\,\cos\left(|\Lam|\,\wt-\frac{\pi(1+d)}{4}\right) ,
\end{eqnarray}
with $\wt=\ln(t+z_m)$. To solve for $\int dt {\cal D}_F(\lam',t)F(\lam,t)=\frac{1}{i}\dlt(\frac{\lam-\lam'}{i})$, we look at
the UV limit of the orthogonality relation of Bessel functions
\begin{align}
&\int_0^\infty xJ_\nu(\xi x)J_\nu(\xi' x)dx=\frac{\dlt(\xi-\xi')}{\xi} \nonumber \\
\Rightarrow \frac{2}{\pi}&\int_0^\infty \cos(\xi x-\frac{\pi\nu}{2}-\frac{\pi}{4})\cos(\xi' x-\frac{\pi\nu}{2}-\frac{\pi}{4})dx
=\dlt(\xi-\xi') \,.
\end{align}
For this case we are able to identify the kernel defined in \eqref{Fredholm_sol} explicitly. 
The expression above implies that ${\cal D}_F(\lam,t)=\frac{\text{sgn}(\Lam)}{|\Lam|^{\frac{1-d}{2}}2i}\cos\left(|\Lam|\wt-\frac{\pi(1+d)}{4}\right)\frac{2}{\pi i}\frac{1+z_m'}{t+z_m}$. This formula is only defined on the imaginary axis, but has
a natural and useful extension in the complex plane,
\begin{equation}
{\cal D}_F(\lam,t)=-\frac{1}{2\pi i}\left(e^{\lam\wt}\frac{1}{K_-(\lam)}-e^{-\lam\wt}\frac{1}{K_+(\lam)}\right)\frac{1+z_m'}{t+z_m} \,.
\end{equation}
To further simplify (\ref{coeff_d}), we define 
\begin{equation}\label{Eq:fbar}
\bar F(\lam,t)=K_+(\lam)(t+z_m)^\lam+K_-(\lam)(t-z_m)^\lam\,,
\end{equation}
 and express
$(t\pm z_m)^\lam$ in terms of $F(\lam,t)$ and $\bar F(\lam,t)$,
\begin{equation}\label{ffbar}
\frac{(t\pm z_m)^\lam}{K_\mp(\lam)}=\frac{\pm F(\lam,t)+\bar F(\lam,t)}{2K_+(\lam)K_-(\lam)}\,.
\end{equation}
Inserting (\ref{ffbar}) into (\ref{coeff_d}), we obtain
\begin{align}\label{split}
\frac{\dlt\phi^d(t)}{\dlt\phi^0(t')}=&\frac{2^{d-1}}{2\pi i}\frac{\Gm(-d)\Gm(\frac{1+d}{2})}{\Gm(\frac{1-d}{2})\Gm(d)}\int dt''d\lam d\lam'  ~\frac{t^{\lam-\frac{1+d}{2}}t'{}^{-\lam'-\frac{1+d}{2}}  }{K_-(\lam')K_+(\lam')}\\ \nonumber
&\times \bigg[P(\lam) {\cal D}_F(\lam,t'')F(\lam',t'') +Q(\lam) {\cal
  D}_F(\lam,t'')\bar F(\lam',t'')\bigg] \, ,
\end{align}
with $P(\lam)=L_+(\lam)K_-(\lam)+L_-(\lam)K_+(\lam)$ and $Q(\lam)=L_+(\lam)K_-(\lam)-L_-(\lam)K_+(\lam)$.
Using the definition of ${\cal D}_F$, the $P(\lam)$ term can be simplified to
\begin{align}
\frac{2^{d-1}}{2\pi i}\frac{\Gm(-d)\Gm(\frac{1+d}{2})}{\Gm(\frac{1-d}{2})\Gm(d)}&\int \left(\frac{L_+(\lam)}{K_+(\lam)}+\frac{L_-(\lam)}{K_-(\lam)}\right)\left(\frac{t}{t'}\right)^{\lam-\frac{1+d}{2}}d\lam \\
&=2^{d-1}\frac{\Gm(\frac{1+d}{2})}{\Gm(\frac{1-d}{2})\Gm(d)}
\left(\theta(t-t')\frac{1}{(t-t')^{d+1}}+\theta(t'-t)\frac{1}{(t'-t)^{d+1}}\right) \,. \nonumber
\end{align}
This is just the average of the retarded and advanced correlators in the vacuum. We will
not include this piece since we are only interested in a state (trajectory) 
dependent contribution to the correlator, which is given by the $Q(\lam)$
term in (\ref{split}). Let us rewrite \eqref{split} without the $P(\lam)$-term in the explicit form
\begin{align}\label{lamp_int}
\frac{\dlt\phi^d(t)}{\dlt\phi^0(t')}=
&-\frac{2^{d-1}}{(2\pi i)^2}\frac{\Gm(-d)\Gm(\frac{1+d}{2})}{\Gm(\frac{1-d}{2})\Gm(d)}\int d\wt d\lam d\lam'  \, \frac{t^{\lam-\frac{1+d}{2}}\,t'{}^{-\lam'-\frac{1+d}{2}} }{K_+(\lam')K_-(\lam')}\left(\frac{e^{\lam\wt}}{K_-(\lam)}-\frac{e^{-\lam\wt}}{K_+(\lam)}\right)  \,\nonumber\\ 
&\times
\left(e^{\lam'\wt}K_+(\lam')+e^{-\lam'\wt}K_-(\lam')\right)\bigg[L_+(\lam)K_-(\lam)
- K_+(\lam)L_-(\lam)\bigg] \, ,
\end{align}
with $K_\pm$, $L_\pm$ given by \eqref{K}, \eqref{LpLm}.
\EPSFIGURE{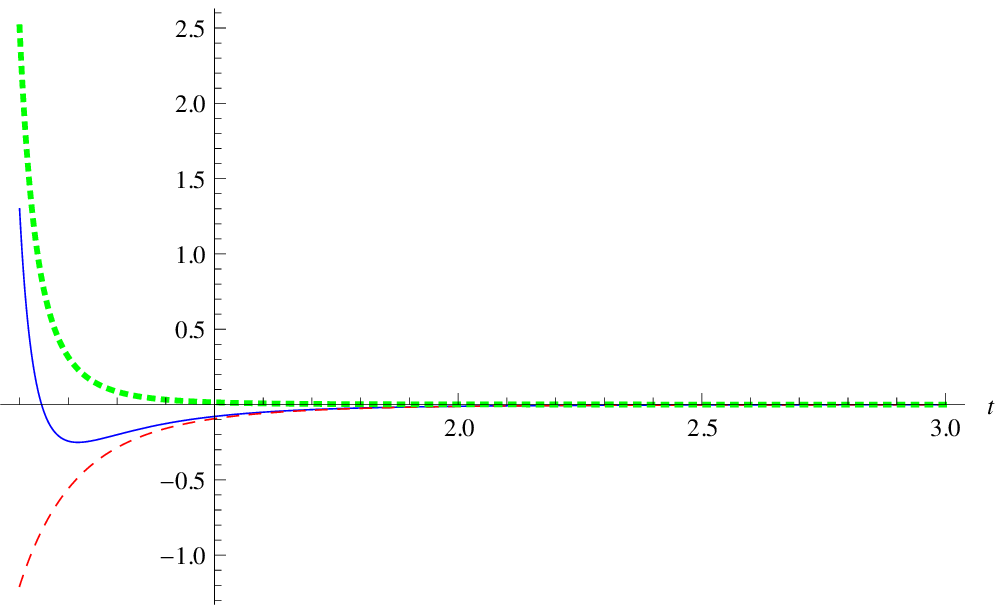,width=10cm}
{
The correlator (\ref{Gttp_hyper}) as a function of $t$ for $t'=2$ at $d=4$. The green dotted and red dashed lines represent the first and second terms in the
square bracket, respectively. The blue solid line is their sum. \label{spacemirror} }
We perform this integral in appendix \ref{AppIntegration}, with the result
\begin{align}\label{Gttp_hyper}
\langle\! \int\! d^{d-1}x~TO(t,x)O(t',0)\rangle=&
~\frac{d~2^{d-1}}{\Gm(d)}\frac{\Gm(\frac{1+d}{2})}{\Gm(\frac{1-d}{2})}
\bigg[-\frac{1}{(tt'-1)^{1+d}} +\frac{\Gm(\frac{1+d}{2})}{2\Gm(\frac{1-d}{2})\Gm(1+d)} \nonumber\\
&\times F\left(1+d,\frac{3+d}{2},2+d;\frac{t'-t}{\left(t'-1\right)t}\right)\frac{1}{(tt'-t)^{1+d}}
\bigg] \, 
\end{align}
for the correlator defined in \eqref{funcderv}. This result is displayed in Fig. \ref{spacemirror}. Note that we have specified the correlator
as a time-ordered one. This is because we focused on only one segment in the 
in-in contour. The causal
nature follows from the principle of the real-time gauge/gravity duality \cite{vanrees}.

We see that the $d$-dependence of the correlator is qualitatively similar to the
constantly moving mirror case. It is again finite for even dimensions and zero for
odd dimensions, a reminiscent of vacuum correlator. 
Turning to the singularities of the correlator, we see that the
first term has singularities at $t=t'=1$. The hypergeometric function in the
second term can be expressed in terms of elementary functions, but this is not 
necessary for our purposes. Singularities appear when the argument $\frac{t'-t}{\left(t'-1\right)t}$ becomes $0,\,1$ or $\infty$, corresponding
to $t=t',\,t=1$ or $t'=1$ respectively. The singularity at $t=t'$ is just the
usual lightcone one. The singularities at $t=1$ and $t'=1$ are closely related
to the starting time of the mirror. No other singularity is found in the
correlator, consistent with the geometric optics expectation. This
example provides  a further
nontrivial check for our approach. 

We note that the relatively simple 
procedure leading to the two-point correlator is related to the $SO(1,1)$ symmetry 
preserved by the ``mirror'' trajectory $t^2-z^2=1$. We expect that a method 
for solving the wave equation exists which is similar 
to that of Section 2, introducing variables which make the symmetry manifest. It will be interesting to see whether the
two-point correlator following from such a method agrees with the result given by the
general prescription, away from the singularities.
\section{Conclusion}

To summarize, we have solved the time-dependent problem of a spatially homogeneous scalar wave equation
in AdS space in the presence of a radially moving mirror. First we considered a
mirror with constant velocity which preserves scaling symmetry.
We used the complete set of solutions to obtain the spatially integrated 
time-dependent two-point function for the CFT state
defined by the moving mirror geometry. We found the result to be  consistent with a
geometric optics picture, in agreement with the bulk-cone
singularities conjecture of  \cite{HLR}.

Moreover, we have determined the precise relation between the 
geometric optics limit and the WKB approximation to the scalar wave solution. The established
connection allows us to solve the scalar wave equation beyond scaling symmetry. As the main
results of this work, we have
established a formula for the two-point function corresponding to a moving mirror geometry
with arbitrary trajectory, valid in the UV limit.  We have tested this formula by reproducing the two-point function
in the case of a mirror moving with constant velocity obtained
before. We performed
a further nontrivial check by considering a spacelike trajectory of
the ``mirror'' (which is not reached by the light ray in this
case). We found that the
singularities of the result are again consistent with a geometric optics picture.

Our analysis provides essential technical tools for the study of two-point
function in time-dependent geometries. These tools may now serve as a
starting point for studying the behaviour of two-point functions
during thermalization in strongly coupled field theories. We describe
some immediate avenues to pursue in the subsequent.

In particular, most immediately
the prescription for the two-point function which we have obtained 
for a moving mirror in AdS space can be generalized
to the case of a gravitational collapsing geometry. 
Specifically, this will involve solving the wave 
equation in a thermal AdS background outside the collapsing shell. This simply 
amounts to a change of the lightcone coordinate from 
$t\pm z$ which we used in the present paper 
to $t\pm \int\frac{dz}{f(z)}$, with $f(z)$ the warp factor 
of the thermal AdS metric. Moreover,
the Dirichlet boundary condition used here has to be replaced by a
 matching condition of a scalar outside and inside the shell. 
The key ingredient of matching the bulk WKB solution 
to the near-boundary eigenmode decomposition (\ref{super}) will remain
unchanged, since the metric of the collapsing shell is asymptotically AdS.
We expect that this  generalization will allow us 
to go beyond the quasi-static limit for the collapsing shell
considered in \cite{quasi} and probe 
the early time dynamics in the thermalization of strongly coupled fields, an important
ingredient in understanding the early thermalization of matter produced in heavy ion collisions \cite{thermalization}.

A further generalization of our results to the boost-invariant background of
\cite{JP} is also possible: The idea is again to match the bulk WKB solution to 
a near-boundary solution, which is expressed as a superposition of the eigenmodes
in pure AdS background.
Due to the additional proper time $\tau$ appearing in the boost invariant metric, 
both the WKB solution and the
eigenmode set consistent with scaling symmetry will take a different form from the
one discussed in this paper. This
will give rise to a proper time dependent correlator with rapidity and transverse space coordinates 
integrated out, instead of a time dependent spatially integrated correlator. A direct application to \cite{JP} with infalling boundary
conditions of the wave imposed on the horizon will allow us to obtain the
quasi-normal modes beyond the
adiabatic approximation used in \cite{JPth}.

Finally, let us compare our results for two-point functions of a
massless scalar field with
light-like separation to previous discussions of space-like separation,
for instance those of
\cite{Banks:1998dd,Balasubramanian:1999zv,ads3}. There it was pointed out that for large
masses, or alternatively for large conformal dimension, the correlator
is dominated by the contribution of the space-like geodesic joining
the two points. Our result is that the pole structure of the correlator
is determined by the geometric optics limit, i.e. by the trajectory of
a light ray, is an analogous statement for the light-like
case valid also for the massless case corresponding to 
small conformal dimension. We expect that our proposed study of
correlators in thermalization geometries along the lines developed in
this paper will also allow for a further study of the role of
geodesics for correlators in these geometries.

\vskip 3cm
\noindent 
{\large {\bf Acknowledgements}}
\vskip .3cm
\noindent 
We are grateful to Ren\'e Meyer and to Felix Rust for collaboration at
early stages of this work. We would like to thank M. Ammon, C. Hoyos,
V. Hubeny, P. Kerner,
P. Kovtun, S. P. Kumar, R. Meyer, A. O'Bannon, A. Sch\"afer, D. Son, D. Teaney and A. Vuorinen for helpful
discussions. We would like to thank the organizers of the ESI program
on ``AdS Holography and the Quark Gluon Plasma'' at the
Erwin-Schr\"odinger-Institute in Vienna and the KITPC program 
``AdS/CFT and Novel Approaches to Hadron and Heavy Ion Physics'' in Beijing
for hospitalities in the course
of this work. S. L.~is supported by the Alexander von Humboldt foundation.
This work is also 
supported in part by the Cluster of Excellence `Origin and Structure
of the Universe'.

\appendix
\section{Evaluation of (3.13)}
The core of the correlator is the following integral
\begin{equation}\label{core}
\frac{1}{2\pi i}\int d\lam \left(\frac{\Gm(\lam+1)}{\Gm(\lam+\frac{1-d}{2})}A+\frac{\Gm(1-\lam)}{\Gm(\frac{1-d}{2}-\lam)}B \right)   \left(\frac{\Gm(\lam+1)}{\Gm(\lam+\frac{1+d}{2})}A+\frac{\Gm(1-\lam)}{\Gm(\frac{1+d}{2}-\lam)}B\right)^{-1}\left(\frac{t}{t'}\right)^\lam .
\end{equation}
\noindent
For definiteness, from \eqref{ratio} we take 
\begin{align}\label{ABExplicit}
A &=(u_0+1)^{\lam}F(\frac{1-d}{2},\frac{1+d}{2};-\lam+1;\frac{1-u_0}{2}) \,,\\ \nonumber  
B &=-(u_0-1)^{\lam}F(\frac{1-d}{2},\frac{1+d}{2};\lam+1;\frac{1-u_0}{2}) \,.
\end{align}
\noindent 
First of all, we note that the integrand (denoted as $F(\lam,t,t')$) has the property:
$F(\lam^*,t,t')=F^*(\lam,t,t')$. Combined with the fact that the integration path
is the imaginary axis, we see the integral is manifestly real, which is
consistent with the reality condition of the correlator.
We will use the residue theorem to evaluate (\ref{core}). The UV part of
the correlator is given by the contribution from poles with large $|\lam|$.
For this, we do an asymptotic expansion of the integrand.
The following properties
of $\Gamma$-functions and Hypergeometric functions are useful \cite{wzx}:
\begin{eqnarray}
\label{App:A} 
\frac{\Gm(\lam+\alpha)}{\Gm(\lam+\beta)}\sim\lam^{\alpha-\beta}\,;\quad \frac{\Gm(-\lam+\alpha)}{\Gm(-\lam+\beta)}\sim
\frac{\sin\pi(\lam-\beta)}{\sin\pi(\lam-\alpha)} \,\lam^{\alpha-\beta} \,;\quad\lim_{\gamma\rightarrow\infty}F(\alpha,\beta;\gamma;z)=1 . 
\end{eqnarray}
A branch cut at the negative real axis is needed, with the argument of $\lam$
fixed by $|arg\, z|<\pi$. Using the above asymptotic behavior, we obtain
the integrand $\sim\lam^d(\frac{t}{t'})^\lam$ as 
$\lam\rightarrow\infty$. If $t>t'$, we close the contour counter-clockwise
and the integral receives contribution from poles in the left half complex
plane, while if $t<t'$, we close the contour clockwise then the integral
receives contribution from poles in the right half complex plane.
\\ \\ \noindent
The possible poles in the whole complex plane are poles of the Gamma function,
Hypergeometric function and roots of 
\begin{equation}\label{root}
\frac{\Gm(\lam+1)}{\Gm(\lam+\frac{1+d}{2})}A+\frac{\Gm(-\lam+1)}{\Gm(-\lam+\frac{1+d}{2})}B=0 \,.
\end{equation}
Note that $F(\alpha,\beta;\gamma,z)$ as a function of $\gamma$
has the same singularities as $\Gm(\gamma)$ \cite{wzx}, we can show that all the
poles of the Gamma function and Hypergeometric function are removable. Thus
we are only left with roots of (\ref{root}).
Due to the non-algebraic nature of 
(\ref{root}), finding analytic expression of all the roots is not possible.
However, we can deduce a general property of the roots: (\ref{root}) can be
equivalently written as
\begin{align}
\frac{\Gm(\lam+1)}{\Gm(\lam+\frac{1+d}{2})} \left(\frac{u_0+1}{u_0-1}\right)^{\frac{\lam}{2}} & F(\frac{1-d}{2},\frac{1+d}{2};-\lam+1;\frac{1-u_0}{2})= \\
&\frac{\Gm(-\lam+1)}{\Gm(-\lam+\frac{1+d}{2})}\left(\frac{u_0+1}{u_0-1}\right)^{\frac{-\lam}{2}}F(\frac{1-d}{2},\frac{1+d}{2};\lam+1;\frac{1-u_0}{2}),\nonumber
\end{align}
or $R(\lam)=R(-\lam)$ with
\begin{equation}
R(\lam)=\frac{\Gm(\lam+1)}{\Gm(\lam+\frac{1+d}{2})}\left(\frac{u_0+1}{u_0-1}\right)^{\frac{\lam}{2}}F(\frac{1-d}{2},\frac{1+d}{2};-\lam+1;\frac{1-u_0}{2})\,.
\end{equation}
We note that $R(\lam^*)=R^*(\lam)$. It is easy to show that if $\lam$ is a root of (\ref{root}), 
$-\lam,\,\lam^*,\,-\lam^*$ are also roots.
We plot the left hand side of (\ref{root}) in the complex $\lam$ plane, and 
find the zeros lie nearly equally spaced on the imaginary axis. Therefore, we conclude that
the roots must be purely imaginary. Now let us determine the asymptotic form of the roots.
 In the limit $\lam\rightarrow\infty\,(\Lambda\rightarrow\infty)$,
(\ref{root}) has the following asymptotic expression
\begin{equation}\label{ABratio}
\lam^{\frac{1-d}{2}}\bigg[(u_0+1)^{\lam}-(u_0-1)^{\lam}e^{\pm i\pi\frac{d-1}{2}}\bigg] \,,
\end{equation}
and the root is given by 
\begin{equation}\label{app_root}
\lam=\pm i\pi\frac{\frac{d-1}{2}+2k}{\ln\frac{u_0+1}{u_0-1}} \,,
\end{equation}
with integer $k\ge 0$. Our approximate roots indeed are consistent with
numerical plots in the sense that they
are symmetric with respect to the real axis and equally spaced. Furthermore 
we expect (\ref{app_root}) to be more accurate when
$\ln\frac{u_0+1}{u_0-1}\rightarrow 0$, i.e. $u_0\rightarrow \infty$. As
the mirror moves more and more slowly, essentially all modes are effectively
UV.

The poles lie precisely along the integration contour of $\lam$. In order to
obtain a well defined result, we have to deform the contour to circumvent
the poles. The ambiguity associated with the detour corresponds
 to the different causal natures of the resulting correlator\footnote{The ambiguity is familiar
 in the standard calculation
of the vacuum correlator. It can be fixed by a prescription of the integration contour
of the frequency}. In practice, it is easy to calculate the advanced correlator, for which we 
shift the integration of $\lam$ slightly to the left. Then
 all the UV poles lie to the right of the contour.
The integration contour has to be closed counter-clockwise, which requires $t'>t$.
In this way, we can avoid the branch cut on the negative real axis.
The residue at each root is obtained with asymptotic expressions as
\begin{equation}
\frac{\frac{\Gm(\lam+1)}{\Gm(\lam+\frac{1-d}{2})}A+\frac{\Gm(-\lam+1)}{\Gm(-\lam+\frac{1-d}{2})}B}{\frac{d}{d\lam}\bigg[\frac{\Gm(\lam+1)}{\Gm(\lam+\frac{1+d}{2})}A+\frac{\Gm(-\lam+1)}{\Gm(-\lam+\frac{1+d}{2})}B\bigg]}\left(\frac{t}{t'}\right)^\lam 
\rightarrow\lam^d\left(\frac{t}{t'}\right)^\lam\frac{1-e^{\mp i\pi d}}{\ln\frac{u_0+1}{u_0-1}} \,.
\end{equation}
We are happy to see the emergence of the factor $(1-e^{\mp i\pi d})$, which
will precisely cancel the pole from $\frac{\Gm(-d)}{\Gm(\frac{1-d}{2})}$. 
Denote $a=\ln\frac{u_0+1}{u_0-1}$ and $b=\ln\frac{t}{t'}$. 
The correlator is given by the sum of residues
\begin{equation}
\langle\! \int\!d^{d-1}xO(t,x)O(t',0)\rangle_A=-d\theta(t'-t)
\sum_{k=0}^{\infty}2^d\lam^de^{b\lam}\frac{1-e^{\mp i\pi d}}{a}\frac{\Gm(-d)\Gm(\frac{1+d}{2})}{\Gm(\frac{1-d}{2})\Gm(d)}
 (tt')^{-\frac{d+1}{2}}\,
\end{equation}
evaluated at $\lam=\pm i\pi\frac{\frac{d-1}{2}+2k}{a}$. The subscript ``A'' stands for the advanced correlator. We find that
the sum over $k$ can be expressed in terms of the Lerch transcendent function 
$\Phi(z,s,\alpha)$
\begin{align}
\langle\! \int\!d^{d-1}xO(t,x)O(t',0)\rangle_A=&-d\sum_{+,-}\theta(t'-t)(tt')^{\frac{-d-1}{2}}
e^{\frac{\pm i(d-1)\pi b}{2a}}(\frac{\pm 2i\pi}{a})^d\frac{1-e^{\mp i\pi d}}{a} \nonumber\\
&\times\frac{\Gm(-d)\Gm(\frac{1+d}{2})}{\Gm(\frac{1-d}{2})\Gm(d)}2^d\Phi(e^{\frac{\pm 2i\pi b}{a}},-d,\frac{d-1}{4}) \,.
\end{align}
As $d\rightarrow\text{integer}$, this reduces to 
\begin{align}
\langle\! \int\!d^{d-1}xO(t,x)O(t',0)\rangle_A=&-d\sum_{+,-}\theta(t'-t)(tt')^{\frac{-d-1}{2}}
\left(\frac{2}{a}\right)^d\frac{1-e^{\mp i\pi d}}{a}\frac{\Gm(-d)\Gm(\frac{1+d}{2})}{\Gm(\frac{1-d}{2})\Gm(d)} \nonumber\\
&\times e^{\pm \frac{i\pi(d-1)}{2}(\frac{b}{a}-c)}
\bigg[\frac{d!}{(-c)^{d+1}(\pm i2\pi)}-\sum_{r=0}^\infty\frac{B_{d+r+1}(\frac{d-1}{4})c^r(\pm i2\pi)^{d+r}}{r!(d+r+1)}\bigg] \,,
\end{align}
where the $B_n(x)$ are the Bernoulli polynomials \cite{table}. The constant $c$ is defined as 
$e^{i2\pi c}=e^{\frac{i2\pi b}{a}}$ with $|c|<\frac{1}{2}$.
To obtain the retarded correlator, we note a useful property of the integrand: $F(\lam,t,t')=F(-\lam,t',t)$.
This leads to the following relation between retarded and advanced correlators,
\begin{equation}
\langle\! \int\!d^{d-1}xO(t,x)O(t',0)\rangle_A=\langle\! \int\!d^{d-1}xO(t',x)O(t,0)\rangle_R \,.
\end{equation}
\section{The Spatially integrated correlator in $d=3$ and $d=4$}
We begin with the unintegrated correlator with conformal dimension $d$. The retarded correlator
is given by \cite{varman}
\begin{equation}
G^R(t,\vec{x})=-i\frac{\Gm(d+1)}{\pi^{d/2}\Gm(\frac{d}{2})}\theta(t)\left(\frac{1}{(-(t-i\eps)^2+\vec{x}^2)^d}-\frac{1}{(-(t+i\eps)^2+\vec{x}^2)^d}\right) \,.
\end{equation}
\noindent
We note the retarded correlator only has support at $t=r=|\vec{x}|$. Integrated with $\int d^{d-1}x$ for $d=3$, we obtain
\begin{equation}
\int d^2xG^R(t,\vec{x})=\frac{6i}{\pi}\left(\frac{1}{(-(t-i\eps)^2+r^2)^2}-\frac{1}{(-(t+i\eps)^2+r^2)^2}\right)\vert^{\infty}_{r=0} \,.
\end{equation}
\noindent
In the limit $\eps\rightarrow 0$, this vanishes identically.
\\ \\ \noindent
For $d=4$, we have, apart from rational function of $t$ and $r$, also the logarithmic function, which makes
the $i\eps$ prescription relevant
\begin{equation}
\int d^3xG^R(t,\vec{x})=-\frac{3}{\pi}(\frac{\ln(r-(t-i\eps))}{(t-i\eps)^5}-\frac{\ln(r-(t+i\eps))}{(t+i\eps)^5})\vert^{\infty}_{r=0}+\cdots
\end{equation}
\noindent
The terms in $\cdots$ drop out as the limit $\eps\rightarrow 0$ is taken. The logarithmic terms give rise to
a finite contribution
\begin{equation}
\int d^3xG^R(t,\vec{x})=-\theta(t)\frac{6}{t^5} \,.
\end{equation}
\noindent
This agrees with $-d\frac{\delta\phi^d(t)}{\delta\phi^0(t')}$ when \eqref{retarded} is used in $d=4$ dimension.
\section{Evaluation of (4.32)}
\label{AppIntegration}
The core part of (\ref{lamp_int}) is given by
\begin{align}\label{app_int}
&\int d\wt d\lam d\lam' \left(\frac{\Gm(\lam+1)\Gm(-\lam+1)}{\Gm(\lam+\frac{1-d}{2})\Gm(\frac{1+d}{2}-\lam)}-\frac{\Gm(\lam+1)\Gm(-\lam+1)}{\Gm(\lam+\frac{1+d}{2})\Gm(\frac{1-d}{2}-\lam)}\right) \frac{\Gm(\lam'+\frac{1+d}{2})\Gm(-\lam'+\frac{1+d}{2})}{\Gm(\lam'+1)\Gm(-\lam'+1)} \nonumber\\
&\times \left(e^{\lam\wt}\frac{\Gm(-\lam+\frac{1+d}{2})}{\Gm(-\lam+1)}-e^{-\lam\wt}\frac{\Gm(\lam+\frac{1+d}{2})}{\Gm(\lam+1)}\right)\left(e^{\lam'\wt}\frac{\Gm(\lam'+1)}{\Gm(\lam'+\frac{1+d}{2})}+e^{-\lam'\wt}\frac{\Gm(-\lam'+1)}{\Gm(-\lam'+\frac{1+d}{2})}\right)
t^{\lam}t'{}^{-\lam'} .
\end{align}
Note that $\wt=\ln(t+\sqrt{t^2-1})\ge 0$. We introduce an upper cutoff $T$ to regularize the $\wt$ integral. With $\wt$ integrated out, (\ref{app_int}) takes the following form
\begin{align}
&\int d\lam d\lam' \left(\frac{\Gm(\lam+1)\Gm(-\lam+1)}{\Gm(\lam+\frac{1-d}{2})\Gm(\frac{1+d}{2}-\lam)}-\frac{\Gm(\lam+1)\Gm(-\lam+1)}{\Gm(\lam+\frac{1+d}{2})\Gm(\frac{1-d}{2}-\lam)}\right)\frac{\Gm(\lam'+\frac{1+d}{2})\Gm(-\lam'+\frac{1+d}{2})}{\Gm(\lam'+1)\Gm(-\lam'+1)} \nonumber\\
&\times \bigg[\frac{\Gm(-\lam+\frac{1+d}{2})\Gm(\lam'+1)}{\Gm(-\lam+1)\Gm(\lam'+\frac{1+d}{2})}\frac{e^{(\lam+\lam')T}-1}{\lam+\lam'} +\frac{\Gm(-\lam+\frac{1+d}{2})\Gm(-\lam'+1)}{\Gm(-\lam+1)\Gm(-\lam'+\frac{1+d}{2})}\frac{e^{(\lam-\lam')T}-1}{\lam-\lam'} \nonumber\\
&~~~~-\frac{\Gm(\lam+\frac{1+d}{2})\Gm(\lam'+1)}{\Gm(\lam+1)\Gm(\lam'+\frac{1+d}{2})}\frac{e^{(-\lam+\lam')T}-1}{-\lam+\lam'} -\frac{\Gm(\lam+\frac{1+d}{2})\Gm(-\lam'+1)}{\Gm(\lam+1)\Gm(-\lam'+\frac{1+d}{2})}\frac{e^{(-\lam-\lam')T}-1}{-\lam-\lam'}
\bigg]t^\lam t'{}^{-\lam'}.
\end{align}
We now use the residue theorem to evaluate the integrals of $\lam$ and $\lam'$.
Note that the appearance of $\frac{e^{(\pm\lam\pm\lam')T}-1}{\pm\lam\pm\lam'}$ does not
introduce any new poles as they have finite limits when $\pm\lam\pm\lam'\rightarrow0$. We begin with the integral of $\lam'$. It is helpful to keep in mind that
$t,\,t'>1$ since the ``mirror'' does not leave the boundary until $t=1$.
Completing the $\lam'$ integral, we find only the first and third terms
in the bracket contribute and 
the dependence on the cutoff $T$ drops out naturally. We obtain the result
\begin{align}\label{app_lam}
&2\pi i  \int  d\lam\left(\frac{\Gm(\lam+1)\Gm(-\lam+1)}{\Gm(\lam+\frac{1-d}{2})\Gm(-\lam+\frac{1+d}{2})}-\frac{\Gm(\lam+1)\Gm(-\lam+1)}{\Gm(\lam+\frac{1+d}{2})\Gm(-\lam+\frac{1-d}{2})}\right) t^\lam  \\
&\times \sum_{n=0}^\infty \frac{(-1)^n}{n!}\frac{1}{\Gm(-n-\frac{1+d}{2}+1)}\bigg[\frac{-t'{}^{-n-\frac{1+d}{2}}}{\lam+n+\frac{1+d}{2}}\frac{\Gm(-\lam+\frac{1+d}{2})}{\Gm(-\lam+1)} +\frac{t'{}^{-n-\frac{1+d}{2}}}{-\lam+n+\frac{1+d}{2}}\frac{\Gm(\lam+\frac{1+d}{2})}{\Gm(\lam+1)}\bigg] \,.\nonumber
\end{align}
We again perform the contour integral. This time it is much simpler: As $t>1$, only the
poles in the left half plane contribute. Considering the first term in the sum,
we find that poles at $\lam+1=-m$ for $m \in \{0,1,2\cdots \}$ and $\lam+n+\frac{1+d}{2}=0$ are
relevant. The contribution from the first set of poles is proportional to
\begin{align}
&\sum_{m=0}^\infty\frac{(-1)^m}{m!}\bigg[\frac{1}{\Gm(-m-1+\frac{1-d}{2})}-\frac{\Gm(m+1+\frac{1+d}{2})}{\Gm(m+1+\frac{1-d}{2})\Gm(-m-1+\frac{1+d}{2})}\bigg]\frac{t^{-m-1}}{m+1-n-\frac{1+d}{2}} \nonumber\\
&=\bigg[\frac{\sin(\frac{\pi(3+d)}{2})\Gm(\frac{3+d}{2})}{\pi}-\frac{1}{\Gm(\frac{-1-d}{2})}\bigg] F\left(\frac{3+d}{2},-n+\frac{1-d}{2};-n+\frac{3-d}{2},\frac{1}{t}\right)\frac{1}{t(n+\frac{d-1}{2})} \,.
\end{align}
We note that the $d$-dependent prefactor vanishes identically by the properties of the
Gamma function. The pole at $\lam+n+\frac{1+d}{2}=0$ gives the contribution
\begin{equation}
\sum_{n=0}^\infty-\frac{(-1)^n}{n!}\left(\frac{1}{\Gm(-n-d)}-\frac{\Gm(1+d+n)}{\Gm(1+n)\Gm(-n)}\right)(tt')^{-n-\frac{1+d}{2}} =-\frac{1}{\Gm(-d)}\frac{(tt')^{\frac{1+d}{2}}}{(tt'-1)^{1+d}} \,,
\end{equation}
where $\frac{1}{\Gm(-d)}$ cancels the overall divergent factor $\Gm(-d)$ to
yield a finite numerical coefficient. Next we consider the second term in the sum of (\ref{app_lam}), the relevant
poles are at $\lam+\frac{1+d}{2}=-m$, with the contributions
\begin{align}
&\sum_{n,m=0}^\infty\frac{(-1)^{m+n}}{m!n!}\frac{\Gm(m+\frac{1+d}{2}+1)}{\Gm(-n-\frac{1+d}{2}+1)\Gm(-m-d)\Gm(m+1+d)}\frac{t^{-m-\frac{1+d}{2}}t'{}^{-n-\frac{1+d}{2}}}{m+n+1+d} \nonumber\\
=&\sum_{n,m=0}^\infty\frac{(-1)^{m+n}}{m!n!}\frac{\sin\pi(n+\frac{d-1}{2})\Gm(n+\frac{1+d}{2})}{\pi}\frac{\sin\pi(m+d)\Gm(m+\frac{1+d}{2}+1)}{\pi}\nonumber\\
&~~~~~~~\times \frac{t'{}^{-n}t^{-m}}{m+n+1+d}(tt')^{-\frac{1+d}{2}}  \nonumber\\
=&\sum_{n,m=0}^\infty\frac{\sin\pi\frac{d-1}{2}\sin\pi d}{\pi^2}\frac{\Gm(\frac{1+d}{2})\Gm(\frac{3+d}{2})}{1+d}\frac{(\frac{1+d}{2})_n(\frac{3+d}{2})_m(1+d)_{m+n}}{(2+d)_{m+n}m!n!}(tt')^{-\frac{1+d}{2}}\,,
\end{align}
where $(\alpha)_m=\frac{\Gm(\alpha+m)}{\Gm(\alpha)}$ is the Pochhammer symbol.
The double sum can be expressed in terms of Appell function, which again can
be converted to a Hypergeometric function
\begin{align}
&\frac{\sin\pi\frac{d-1}{2}\sin\pi d}{\pi^2}\frac{\Gm(\frac{1+d}{2})\Gm(\frac{3+d}{2})}{1+d}F_1\left(1+d,\frac{3+d}{2},\frac{1+d}{2},2+d;\frac{1}{t},\frac{1}{t'}\right)(tt')^{-\frac{1+d}{2}} \\
=&\frac{\sin\pi\frac{d-1}{2}\sin\pi d}{\pi^2}\frac{\Gm(\frac{1+d}{2})\Gm(\frac{3+d}{2})}{1+d}\left(1-\frac{1}{t'}\right)^{-1-d}F\left(1+d,\frac{3+d}{2},2+d;\frac{t'-t}{\left(t'-1\right)t}\right)(tt')^{-\frac{1+d}{2}} \,. \nonumber
\end{align}
Writting $\sin$-functions as product of two $\Gamma$-functions, collecting all nonvanishing
terms and inserting the overall factor, we end up with the following result
\begin{align}
\frac{\dlt\phi^d(t)}{\dlt\phi^0(t')}=&-\frac{\Gm(\frac{1+d}{2})2^{d-1}}{\Gm(\frac{1-d}{2})\Gm(d)}
\bigg[-\frac{1}{(tt'-1)^{1+d}} \\
&+\frac{\Gm(\frac{1+d}{2})}{2\Gm(\frac{1-d}{2})\Gm(1+d)} F\left(1+d,\frac{3+d}{2},2+d;\frac{t'-t}{\left(t'-1\right)t}\right)\frac{1}{(tt'-t)^{1+d}}
\bigg] \, .   \nonumber
\end{align}

\end{document}